\documentclass[aps,prd,twocolumn,nofootinbib,showkeys,preprintnumbers,reprint,floatfix]{revtex4-1}
\usepackage{amsmath}
\usepackage{amssymb}
\usepackage{graphicx}
 \usepackage{color}
\usepackage[T1]{fontenc} 
\usepackage{slashed}
\usepackage{ulem}

\newcommand{\Br}{\text{BR}}

\definecolor{mkgreen}{rgb}{0.2,.70,.3}

\newcommand{\MSUSY}{M_{SUSY}}
\newcommand{\EQ}[1]{eq.~(\ref{#1})}

\def\gsim{\raise0.3ex\hbox{$\;>$\kern-0.75em\raise-1.1ex\hbox{$\sim\;$}}}

\newcommand{\AddrBonn}{%
Physikalisches Institut der Universit\"at Bonn, 53115 Bonn, Germany
}

\newcommand{\AddrOrsay}{%
Laboratoire de Physique Th\'eorique, CNRS -- UMR 8627, 
Universit\'e de Paris-Sud 11\\ F-91405 Orsay Cedex, France}

\newcommand{\AddrWue}{%
Institut f\"ur Theoretische Physik und Astronomie, 
Universit\"at W\"urzburg\\
97074 W\"urzburg, Germany}

\newcommand{\AddrLiege}{%
IFPA, Dep. AGO, Universit\'e de Li\`ege, Bat B5, Sart-Tilman B-4000
Li\`ege 1, Belgium 
}

\newcommand{\AddrMadrid}{%
Departamento de F\'isica Te\'orica, Universidad Aut\'onoma de Madrid,
Cantoblanco, Madrid 28049, Spain
}

\newcommand{\AddrMadridb}{%
Instituto de F\'isica Te\'orica UAM/CSIC,
Calle Nicol\'as Cabrera 13-15, Cantoblanco, Madrid 28049, Spain
}

\begin{document}

\preprint{Bonn-TH-2013-09}
\preprint{LPT-Orsay-13-140}
\preprint{FTUAM-14-39}
\preprint{IFT-UAM/CSIC-14-065 }


\title{Decoupling of heavy sneutrinos in low-scale seesaw models}

\author{M. E. Krauss} \email{manuel.krauss@physik.uni-wuerzburg.de}
%\cortext[cor1]{Corresponding author}

\author{W. Porod} \email{porod@physik.uni-wuerzburg.de}
\affiliation{\AddrWue}

\author{F. Staub} \email{fnstaub@physik.uni-bonn.de}
\affiliation{\AddrBonn}

\author{A. Abada} \email{asmaa.abada@th.u-psud.fr}
\affiliation{\AddrOrsay}

\author{A. Vicente} \email{avelino.vicente@ulg.ac.be}
\affiliation{\AddrLiege}

\author{C. Weiland} \email{cedric.weiland@uam.es}
\affiliation{\AddrOrsay} \affiliation{\AddrMadrid} \affiliation{\AddrMadridb}

%\address[wu]{\AddrWue}
%\address[bonn]{\AddrBonn}
%\address[ors]{\AddrOrsay}
%\address[liege]{\AddrLiege}
%\address[madrid]{\AddrMadrid}
%\address[madridb]{\AddrMadridb}
%\fntext[fn]{Present address: \AddrMadrid;\\ \AddrMadridb}
\begin{abstract}
There have been some recent claims in the literature about large
right-handed sneutrinos contributions to lepton flavor violating
observables like $\mu \to 3e$ or $\mu - e$ conversion in nuclei in supersymmetric 
low-scale seesaw models. These large contributions originate from
$Z$-penguin diagrams which show a much weaker dependence on the heavy
masses than the photonic contributions.  We have traced this to an
error in the evaluation of the corresponding loop amplitudes which has
propagated in the literature. We explicitly show that after correcting
this mistake the $Z$-penguins show the expected decoupling
behavior. Moreover, the reported dominance of the $Z$-penguin over
the photonic contributions disappears as well.\\
%\newline
%\textit{Keywords:} lepton flavor violation, supersymmetry, decoupling, seesaw mechanism
\end{abstract}

\maketitle

\section{Introduction}
\label{sec:intro}

Flavor violation in the neutrino sector is nowadays a well-established fact \cite{Beringer:1900zz}. The mixing angles and the
squared mass differences have been measured with increasing precision
in the last year
\cite{Tortola:2012te,GonzalezGarcia:2012sz,Capozzi:2013csa}.  Lepton
flavor violation (LFV) in the neutrino sector implies of course also
LFV in the charged lepton sector. However, in this case only severe
upper limits on LFV branching ratios, such as those for $\mu \to e
\gamma$ \cite{Adam:2013mnn} or $\mu \to 3 e$ \cite{Bellgardt:1987du},
exist. Many neutrino mass models typically predict sizeable and well
measurable effects in this sector.  Widely studied examples are
supersymmetric variants of high-scale seesaw models
\cite{Minkowski:1977sc,Yanagida:1979as,GellMann:1980vs,Schechter:1980gr,Cheng:1980qt,Foot:1988aq},
see,
e.g.~\cite{Hisano:1995cp,Deppisch:2002vz,Arganda:2005ji,Petcov:2005yh,Antusch:2006vw,Paradisi:2006jp,Hirsch:2008dy,Hirsch:2008gh,%
  Biggio:2010me,Esteves:2010ff,Abada:2010kj,Abada:2011mg,Hirsch:2012yv,Cannoni:2013gq}. In
these kinds of models, the flavor violation in the neutral sector is
transmitted to the charged sector in the renormalization group
evolution from the high scale where the supersymmetry (SUSY) breaking parameters are specified to the low scale \cite{Borzumati:1986qx}.
A generic prediction for the radiative decay $\ell_j \to \ell_i \gamma$
reads~\cite{Hisano:1995cp}
\begin{equation}
 \Br(\ell_j \to \ell_i \gamma) \simeq \frac{48 \pi^3 \alpha}{G_F^2}
\frac{|(m^2_{\tilde f})_{ji}|^2}{\MSUSY^8} \Br(\ell_j \to \ell_i \nu_j{\bar\nu_i}) \, .
\end{equation}
Here $(m^2_{\tilde f})_{ji}$ parametrizes the largest off-diagonal
elements of the soft-breaking slepton mass matrices and $\MSUSY$ is
the typical mass of the supersymmetric particles, nowadays expected to
lie in the TeV range. If one does not rely on special cancellations,
usually small off-diagonal elements are required to satisfy
experimental bounds \cite{Beringer:1900zz}. Since in high-scale seesaw
models the photonic penguin contributions dominate also the decay $\ell_j
\to 3\ell_i$ a simple relation between both observables exists
\cite{Ilakovac:1994kj,Arganda:2005ji}
\begin{equation}
\label{eq:relation}
\Br(\ell_j \to 3 \ell_i) \simeq \frac{\alpha}{3 \pi} 
\left(\log\left(\frac{m^2_{\ell_j}}{m^2_{\ell_i}}\right) - \frac{11}{4} \right) 
\Br(\ell_j \to \ell_i \gamma)\,.
\end{equation}
Therefore, the radiative decay $\ell_j \to \ell_i \gamma$ is in general more
constraining than $\ell_j \to 3 \ell_i$ (up to some exceptions
\cite{Babu:2002et,Paradisi:2005tk}).

Recently, low-scale seesaw scenarios (such as the inverse seesaw) have
gained more interest.  In the inverse seesaw~\cite{Mohapatra:1986bd},
the minimal supersymmetric standard model (MSSM) particle content is extended by three generations of
right-handed neutrino superfields $\hat{\nu}^c$ and of gauge singlets
$\hat{N}_S$, which carry lepton number. The superpotential reads
\begin{equation}
\label{eq:W_lin_inv}
 W_{IS} = W_{\text{MSSM}} + Y_{\nu}\hat{\nu}^c\hat{L}\hat{H}_u
+M_R\,\hat{\nu}^c{\hat{N}}_{S}
+ \frac{\mu_N}{2}{\hat{N}}_{S}{\hat{N}}_{S} \,.\\
\end{equation}
After electroweak symmetry breaking (EWSB), the effective mass matrix
for the light neutrinos is approximately given by $m_\nu \simeq
\frac{v_u^2}{2} Y_\nu (M^T_R)^{-1} \mu_N M^{-1}_R Y^T_\nu$. The
additional suppression given by $\mu_N$ allows for Yukawa couplings of
order $\mathcal{O}(1)$ and $M_R \sim \mathcal{O}(\MSUSY)$ while being
compatible with neutrino oscillation data.

In Ref.~\cite{Hirsch:2012ax}, the relation in \EQ{eq:relation} was
found to be violated in the inverse seesaw due to a large enhancement
of the $Z$-penguins. These contributions, enhanced with respect to
photonic penguins by a factor $\left(\MSUSY^4/m_Z^4\right)$, turned
out to be dominant in most parts of parameter space, specially in case
of large $M_R$. 
% changed:
Later, this qualitative result was further exploited in several
phenomenological studies \cite{Dreiner:2012mx,Hirsch:2012kv,Abada:2012cq}\footnote{In an independent calculation \cite{Ilakovac:2012sh}, a $Z$-penguin dominance was found for a specific choice of mSUGRA parameters.}.
Furthermore, this
enhancement in the $Z$-penguins was interpreted sometimes as a
nondecoupling behavior. This nondecoupling behavior would imply the
existence of a flavor violating $Z \ell_i \ell_j$ operator without any
suppression from the new physics scale $\Lambda$.  In an expansion in
powers of $\frac{v}{\Lambda}$, where $v$ is the electroweak VEV, this
operator would contain a nonvanishing zero order contribution. This
is, however, absent in well-known lists of allowed effective operators
\cite{Buchmuller:1985jz} as it would introduce an explicit violation
of the SM gauge symmetry.  Therefore a critical discussion of this
effect becomes necessary.

While most previous works
\cite{Hirsch:2012ax,Dreiner:2012mx,Hirsch:2012kv,Abada:2012cq} have
adapted well established results of the MSSM \cite{Arganda:2005ji}, we
perform here a completely independent calculation of the $Z$-penguin
contributions to $\ell_j\to 3\ell_i$ and $\mu-e$ conversion. 
% changed
We find that
the $Z$-penguins do not show the dominant behavior reported
in~\cite{Hirsch:2012ax,Abada:2012cq}. The reason stems from a
mistake in the $Z$-penguin contributions already present in the MSSM
results of Ref.~\cite{Arganda:2005ji}. While the mistake
in the prediction of charged LFV rates has no impact in the case of high-scale
seesaw models, for low-scale seesaw models it changes the
entire picture.

We present in the next section our revised calculation of the
$Z$-penguin, which contributes to several LFV observables, and point out the
difference to previous calculations in the literature. 
%In sec.~\ref{sec:numerics} 
Afterward we numerically compare the old and new results
before we conclude.
% in sec.~\ref{sec:conclusion}
In the Appendix we provide the vertices and loop functions that are used in the computation.
\newline

\section{Revising the 1-loop $\ell_i - \ell_j - Z$ effective coupling}
\label{sec:analytical}

LFV violating observables have been intensively discussed in
supersymmetric high-scale seesaw models
\cite{Hisano:1995cp,Deppisch:2002vz,Arganda:2005ji,Petcov:2005yh,Antusch:2006vw,Paradisi:2006jp,Hirsch:2008dy,Hirsch:2008gh,%
  Biggio:2010me,Esteves:2010ff,Abada:2011mg,Hirsch:2012yv,Cannoni:2013gq}. In
view of the above-mentioned controversy, we focus on the $Z$-penguin
and, in particular, on the chargino-sneutrino contributions. We
consider the definition of the effective $\ell_i - \ell_j - Z$ vertex
%\begin{align}
%\bar u_i(p_1) \gamma_\mu \big(F_L P_L + F_R P_R\big) u_j(p_2) \epsilon^\mu(p_3)
%\end{align}
\begin{align}
\bar \ell_j \gamma_\mu \big(F_L P_L + F_R P_R\big) \ell_i Z^\mu \, .
\end{align}
The form factors $F_{L,R}$ contribute to several LFV processes, among
which one can find $\ell_j \to 3 \ell_i$ \cite{Arganda:2005ji}, $\mu-e$
conversion in nuclei \cite{Arganda:2007jw} and $\tau$ mesonic LFV
decays \cite{Arganda:2008jj}.  In Ref.~\cite{Arganda:2005ji}, the
chargino contributions to the form factor $F_L$ are found to be
\begin{widetext}
\begin{align}
\label{eq:AH}
\notag F_L^{(c)} = -\frac{1}{16 \pi^2} &\Big(
C^R_{iBX} C^{R*}_{jAX} \big( 2 E^{R(c)}_{BA} C_{24} (m^2_{\tilde \nu_X},m^2_{\tilde \chi_A^-},m^2_{\tilde \chi_B^-}) - 
 E^{L(c)}_{BA} m_{\tilde \chi_A^-} m_{\tilde \chi_B^-} C_0(m^2_{\tilde \nu_X},m^2_{\tilde \chi_A^-},m^2_{\tilde \chi_B^-})\big) \\ 
 &+C^R_{iAX} C^{R*}_{jAY} \big( 2 Q^{\tilde \nu}_{XY} C_{24} (m^2_{\tilde \chi^-_{A}},m^2_{\tilde \nu_X},m^2_{\tilde \nu_Y}) \big)  
 + C^R_{iAX} C^{R*}_{jAX} Z^{(\ell)}_L B_1(m^2_{\tilde \chi_A^-},m^2_{\tilde \nu_X})
\Big)\,,
\end{align}
\end{widetext}
where $C^R_{iAX}$, $E^{R(c),L(c)}_{BA}$, $Q^{\tilde \nu}_{XY}$ and $Z^{(\ell)}_L$ are the 
$\tilde \chi_A-\ell_i-\tilde \nu_X$, $\tilde \chi^A -\tilde \chi^B - Z$, 
 $\tilde \nu_X - \tilde \nu_Y -Z$ and $\ell-\ell-Z$ couplings, respectively. For detailed 
definitions see Appendix \ref{appendix:vertices} or \cite{Arganda:2005ji}. $C_0$, $B_1$ and $C_{24}$ are loop 
functions evaluated in the limit of zero external momenta. $C_0$ and $B_1$ are well-known Passarino-Veltman functions, 
whereas combining the definitions in \cite{Arganda:2005ji} and  \cite{Arganda:2004bz}
$C_{24}$ is given by
\begin{align}
\label{EQ:C24}
4 \, C_{24}(m_0^2,m_1^2,m_2^2) = B_0(m_1^2,m_2^2) + m_0^2 C_0(m_0^2,m_1^2,m_2^2)\,.
\end{align}

It proves convenient to expand $F_L^{(c)}$ in powers of the chargino
mixing angle. This allows one to get a clear understanding of the
leading contributions. The lowest order in the expansion corresponds
to zero chargino mixing, which we further split as $F_L^{(c,0)} =
-\frac{1}{16 \pi^2}\big( \mathcal F^{\tilde W(0)}_{L} + \mathcal
F^{\tilde H(0)}_{L} \big)$. Here $\mathcal F^{\tilde W(0)}_{L}$
represents the pure wino contribution and $\mathcal F^{\tilde
  H(0)}_{L}$ the pure Higgsino contribution.  As pointed out in
Ref.~\cite{Hirsch:2012ax}, using this method (and the results for the
loop functions in \cite{Arganda:2004bz,Arganda:2005ji}) one can show
that the contribution for a pure wino and a pure left-handed sneutrino
vanishes exactly in the MSSM.  These equations can be easily adjusted
to the inverse seesaw~\cite{Abada:2012cq}. In this case, the contribution
from the pure wino and pure left-handed sneutrino vanishes again, as
in the MSSM. However, one finds a nonzero contribution from pure
Higgsino and pure right-handed sneutrino loops
\begin{equation}
\label{EQ:abada}
\mathcal F^{\tilde H (0)}_{L} = \frac{g}{8 \cos \theta_W}\big(Y_\nu^\dagger Y_\nu \big)_{ij} \Big( \cos^2\theta_W -\frac12 \Big) \, .
\end{equation}

The result in \EQ{EQ:abada} does not depend on the SUSY scale (nor on
$M_R$), which clearly looks like a nondecoupling effect. However, we
will show that it is indeed caused by an error in \EQ{eq:AH}. We
recalculated the chargino contributions and found, in contrast to
\EQ{eq:AH}, the 1-loop $Z - \ell_i - \ell_j$ effective coupling

\begin{widetext}
\begin{align}
\notag F_L^{(c)} = -\frac{1}{16 \pi^2} &\Big(
C^R_{iBX} C^{R*}_{jAX} \big( E^{R(c)}_{BA} \big[B_0(m^2_{\tilde \chi_A^-},m^2_{\tilde \chi_B^-})- 2 C_{00} (m^2_{\tilde \nu_X},m^2_{\tilde \chi_A^-},m^2_{\tilde \chi_B^-}) +    m^2_{\tilde \nu_X} C_0(m^2_{\tilde \nu_X},m^2_{\tilde \chi_A^-},m^2_{\tilde \chi_B^-})\big] \\ \notag
 &-E^{L(c)}_{BA} m_{\tilde \chi_A^-} m_{\tilde \chi_B^-} C_0(m^2_{\tilde \nu_X},m^2_{\tilde \chi_A^-},m^2_{\tilde \chi_B^-})\big) +
  C^R_{iAX} C^{R*}_{jAY} \big( 2 Q^{\tilde \nu}_{XY} C_{00} (m^2_{\tilde \chi^-_{A}},m^2_{\tilde \nu_X},m^2_{\tilde \nu_Y}) \big)  \\ 
 &+C^R_{iAX} C^{R*}_{jAX} Z^{(\ell)}_L B_1(m^2_{\tilde \chi_A^-},m^2_{\tilde \nu_X})
\Big)\,.
\label{EQ:us}
\end{align}
\end{widetext}

We must now compare this result to \EQ{eq:AH} by using the relation
between the loop functions in the limit of zero external momenta
squared \cite{Eberl:diss},
\begin{align}
\label{EQ:eberl}
D \, C_{00} (m^2_0,m^2_1,m^2_2) = B_0(m^2_1,m^2_2) + m_0^2 C_0 (m^2_0,m^2_1,m^2_2)\,.
\end{align}
where $D=4-2 \epsilon$ in dimensional regularization/reduction.  With
these definitions, we can relate the expression of $C_{24}$ in Eq. (\ref{EQ:C24}) to $C_{00}$
via $C_{24} = C_{00}-\frac18$ since $D C_{00} = 4 C_{00} -
\frac12$. We find that eqs. \eqref{eq:AH} and \eqref{EQ:us} would agree
if we (incorrectly) used $D C_{00} = 4 C_{00}$ and identified $C_{24}$ with $C_{00}$. This makes obvious
that our results differ by constant terms which seem to originate from
the handling of $1/\epsilon$ singularities in the loop calculation. In
the following, we will show how these differences impact the
decoupling behavior by explicitly recalculating \EQ{EQ:abada}, and
showing that it indeed vanishes.
  
As a technical detail we note that the Majorana mass terms in the
neutrino sector also induce a splitting of the sneutrinos into their
scalar and pseudoscalar components. While this splitting has to be
tiny for left sneutrinos \cite{Hirsch:1997vz,Grossman:1997is} it can
be quite sizable for the gauge singlets.  As this can lead in
principle to visible effects we include it in the following
discussion.  The part of the effective coupling $F_L^{(c)}$ that is
proportional to $Y_\nu^\dagger Y_\nu$, and thus projects onto the
Higgsino in the loop, reads

\begin{widetext}
\begin{align}
 \notag \mathcal F_L^{\tilde H} =& -\frac{1}{4}\sum_{P,S} 
                 Y_{\nu,ai}^* Y_{\nu,bj}  V_{B2} V^*_{A2} 
                  \big (\mathcal A^\text{wave}_{abAB} + \mathcal A^\chi_{abAB} + \mathcal A^\nu_{abAB} \big)\,, \\ \notag
        \mathcal A^\text{wave}_{abAB} =&  - Z^{P/S *}_{X,3+a} Z^{P/S *}_{X,3+b}\delta_{BA} \big(g_2 \cos{\theta_W} - g_1 \sin{\theta_W} \big) 
         B_1(m^2_{\tilde \chi^-_A},m^2_{\tilde \nu_X}) \,,    \\ \notag
        \mathcal A^\chi_{abAB} =& Z^{P/S *}_{X,3+a} Z^{P/S *}_{X,3+b}\Big[\Big(2 g_2  \cos\theta_W V_{B1}^* V_{A1}  + 
          V_{B2}^* V_{A2} \big(g_2 \cos\theta_W - g_1 \sin\theta_W \big) \Big)\times \\ \notag  &
                 \Big(2 C_{00} (m^2_{\tilde \nu_X},m^2_{\tilde \chi^-_A},m^2_{\tilde \chi^-_B}) - B_0(m^2_{\tilde \chi^-_A},m^2_{\tilde \chi^-_B}) -
                 m_{\tilde \nu_X}^2 C_0(m^2_{\tilde \nu_X},m^2_{\tilde \chi^-_A},m^2_{\tilde \chi^-_B}) \Big) +  \\ \notag 
    & \Big((2 g_2  \cos\theta_W U_{A1}^* U_{B1} +  U_{A2}^* U_{B2} (g_2 \cos\theta_W - g_1 \sin\theta_W )\Big) %\times \\ \notag  &
                 m_{\tilde \chi_A} m_{\tilde \chi_B}  C_0(m^2_{\tilde \nu_X},m^2_{\tilde \chi^-_A},m^2_{\tilde \chi^-_B}) \Big]\,,\\ 
        \mathcal A^\nu_{abAB} =&\big(g_2 \cos\theta_W + g_1 \sin\theta_W \big) \delta_{BA}  Z^{P/S*}_{Xc} Z^{S/P*}_{Yc}  Z^{P/S *}_{X,3+a} Z^{S/P*}_{Y,3+b}
        2 C_{00}(m^2_{\tilde \chi^-_A},m^2_{\tilde \nu_X},m^2_{\tilde \nu_Y})\,.
\end{align}
\end{widetext}
$Z^{P/S}$ represent the mixing matrix of the (pseudo)scalar
sneutrinos. $U$ and $V$ are the usual unitary matrices that
diagonalize the chargino matrix, with the $k1~(k2)$ component
projecting on the Wino (Higgsino) component of $\tilde
\chi^\pm_k$. The sneutrino mass $m_{\tilde \nu_k}$ corresponds to the
respective CP-state, with the index $k$ covering all mass
eigenstates. Sums over repeated indices are implicitly understood and
$a,b,c = 1,\dots,3$.  In the limit of zero chargino mixing, i.e. for
$V$ and $U$ being identity matrices, the expression simplifies to
\begin{align}
 \notag \mathcal F_L^{\tilde H (0)} &= -\frac{1}{4}\sum_{P,S}  
                 Y_{\nu,ai}^* Y_{\nu,bj} 
                 \big(g_2 \cos\theta_W - g_1 \sin\theta_W \big) \mathcal A^\text{sum}_{ab}\,,\\ \notag
         \mathcal  A^\text{sum}_{ab} &= Z^{P/S *}_{X,3+a} Z^{P/S *}_{X,3+b} \Big(- B_1(m^2_{\tilde \chi^-_2},m^2_{\tilde \nu_X})   +\\ \notag 
        &(m^2_{\tilde \chi^-_2} - m_{\tilde \nu_X}^2 ) C_0(m^2_{\tilde \nu_X},m^2_{\tilde \chi^-_2},m^2_{\tilde \chi^-_2}) +\\ \notag
                 &2 C_{00} (m^2_{\tilde \nu_X},m^2_{\tilde \chi^-_2},m^2_{\tilde \chi^-_2}) - B_0(m^2_{\tilde \chi^-_2},m^2_{\tilde \chi^-_2})\Big) + \\ 
                 & 2 Z^{P/S*}_{Xc} Z^{S/P*}_{Yc} Z^{P/S *}_{X,3+a} Z^{S/P *}_{Y,3+b} 
                  C_{00}(m^2_{\tilde \chi^-_2},m^2_{\tilde \nu_X},m^2_{\tilde \nu_Y})\,.
\end{align}
If the left and right sneutrinos do not mix among each other, $\mathcal A^\text{sum}$ reduces to 
\begin{align}
\notag  \mathcal A^\text{sum}_{ab} =& Z^{P/S *}_{X,3+a} Z^{P/S *}_{X,3+b} \Big( - B_1(m^2_{\tilde \chi^-_2},m^2_{\tilde \nu_X}) +\\ \notag  &  (m^2_{\tilde \chi^-_2} - m_{\tilde \nu_X}^2 ) C_0(m^2_{\tilde \nu_X},m^2_{\tilde \chi^-_2},m^2_{\tilde \chi^-_2}) + \\  &
                 2 C_{00} (m^2_{\tilde \nu_X},m^2_{\tilde \chi^-_2},m^2_{\tilde \chi^-_2}) - B_0(m^2_{\tilde \chi^-_2},m^2_{\tilde \chi^-_2})\Big)\,.
\end{align}
Using the explicit expressions for the loop functions (see,
e.g. \cite{delAguila:2008zu}), one can immediately see that the term
in the brackets vanishes.  We can compare this expression with the
nonvanishing one of Ref. \cite{Abada:2012cq} by again using
\EQ{EQ:eberl}. One obtains

\begin{align}
\notag \mathcal A^{\text{sum} \, \prime}_{ab} &=Z^{P/S *}_{X,3+a} Z^{P/S *}_{X,3+b} \Big( - B_1(m^2_{\tilde \chi^-_2},m^2_{\tilde \nu_X})+  \\ \notag  &m^2_{\tilde \chi^-_2} C_0(m^2_{\tilde \nu_X},m^2_{\tilde \chi^-_2},m^2_{\tilde \chi^-_2}) -
                 2 C_{00} (m^2_{\tilde \nu_X},m^2_{\tilde \chi^-_2},m^2_{\tilde \chi^-_2}) \\
                 &+ \frac12 \Big)\,.
\end{align}

Our result differs by a mass independent term of $\frac14$ from the results of Ref. \citep{Abada:2012cq}, which leads to the disappearance of the nondecoupling contribution\footnote{Note that an additional
 different overall factor of $\frac12$ can be traced back to the part $Z^{P/S *}_{X,3+a} Z^{P/S *}_{X,3+b} = \delta_{ba}$ which was wrongly taken to be $\frac12 \delta_{ba}$ in \citep{Abada:2012cq}.}.

Finally, the results for the pure wino contribution read
\begin{align}
  \mathcal F_L^{\tilde W (0)} =& -\frac{1}{4}\sum_{P,S}  
                 g_2^2 \big(g_2 \cos \theta_W Y_1 + g_1 \sin \theta_W Y_2 \big)\,,\\ \notag
        Y_1 =& Z^{P/S *}_{Xi} Z^{P/S *}_{Xj} \Big( - B_1(m^2_{\tilde \chi^-_1},m^2_{\tilde \nu_X})   + \\ \notag &2 (m^2_{\tilde \chi^-_1} - m_{\tilde \nu_X}^2 ) C_0(m^2_{\tilde \nu_X},m^2_{\tilde \chi^-_1},m^2_{\tilde \chi^-_1}) + \\ \notag &
                 4 C_{00} (m^2_{\tilde \nu_X},m^2_{\tilde \chi^-_1},m^2_{\tilde \chi^-_1}) -  2 B_0(m^2_{\tilde \chi^-_1},m^2_{\tilde \chi^-_1})\Big) + \\ \notag
                 & 2  Z^{P/S*}_{Xc} Z^{S/P*}_{Yc} Z^{P/S *}_{Xi} Z^{S/P *}_{Yj}
                 C_{00}(m^2_{\tilde \chi^-_1},m^2_{\tilde \nu_X},m^2_{\tilde \nu_Y})\,, \\ \notag
        Y_2 =& Z^{P/S *}_{Xi} Z^{P/S *}_{Xj} B_1(m^2_{\tilde \chi^-_1},m^2_{\tilde \nu_X}) + \\ \notag &2  Z^{P/S*}_{Xc} Z^{S/P*}_{Yc} Z^{P/S *}_{Xi} Z^{S/P *}_{Yj}
                 C_{00}(m^2_{\tilde \chi^-_1},m^2_{\tilde \nu_X},m^2_{\tilde \nu_Y})\,.
\end{align}
$Y_1$ and $Y_2$ both vanish if there is no left-right mixing among the
sneutrinos and no mass splitting of the CP-even and CP-odd sneutrino
states (i.e., in the MSSM limit). Notice that $Y_1$ and $Y_2$ vanish
because of an exact cancellation of the two combinations of loop
functions. In contrast, the expressions in Refs. \cite{Abada:2012cq,
  Hirsch:2012ax} only vanish for flavor changing transitions (due to
the unitarity of the sneutrino mixing matrix), but still contain
nonzero diagonal entries in the MSSM limit, $Y_1 \to -\frac34
\delta_{ij}$ and $Y_2 \to -\frac14 \delta_{ij}$.
$Y_1$ again differs by a constant term, analogously to $\mathcal
A^\text{sum}$.  $Y_2$ is the same as in \citep{Abada:2012cq} but
vanishes due to the
usage of $C_{00}$ instead of $C_{24}$. Therefore, although the conclusion is the same, the
cancellations in the off-diagonal wino contributions have different
origins. Chargino mixing, of course, spoils all of these cancellations
and $Y_i$ depend on the details of the sneutrino mixing matrix such as
left-right mixing, left-left mixing as well as a mass splitting of the
CP-even and CP-odd sneutrinos. Note that the mass splitting of the
distinct CP eigenstates could in principle give large effects in the
amplitudes. However, in practice it can safely be neglected since it
is tightly constrained by neutrino data to be very small
\cite{Hirsch:1997vz,Grossman:1997is}.

\section{Numerical results}
\label{sec:numerics}

For the numerical study of our new analytical results, we have created
a {\tt SPheno} \cite{Porod:2003um,Porod:2011nf} version for the
inverse seesaw using a modified version of {\tt SARAH}
\cite{Staub:2008uz,Staub:2009bi,Staub:2010jh,Staub:2012pb,Staub:2013tta}. We
parametrize the Yukawa couplings $Y_\nu$ according
to~\cite{DeRomeri:2012qd}:
\begin{align}
\notag  Y_\nu =& f \begin{pmatrix}
          0 & 0 & 0 \\
          a & a (1-\frac{\sin \theta_{13}}{\sqrt{2}}) & - a (1+\frac{\sin \theta_{13}}{\sqrt{2}}) \\
          \sqrt{2} \sin \theta_{13} & 1 & 1
         \end{pmatrix} \label{eq:yv_parametrization}, \\
    &a = \left(\frac{\Delta m_{\odot}^2}{\Delta m_\text{Atm}^2}\right)^{\frac{1}{4}} \approx 0.4\, ,
\end{align}
using the data from the global fit of the Pontecorvo-Maki-Nakagawa-Sakata (PMNS) matrix
\cite{GonzalezGarcia:2012sz}\footnote{Note that this parametrization is not general but merely corresponds to one possibility in which $\mu_N$ is diagonal and the lightest neutrino eigenstate massless.}.

In the following discussion we set for the sake of illustration all the slepton soft SUSY breaking
masses to $\MSUSY$. The soft SUSY breaking gaugino masses are
scaled as $\MSUSY$ and their starting values are $M_1=80$~GeV and
$M_2=160$~GeV. The $A$-parameters in the slepton sector are set to
130 GeV. 
Moreover, we set $\tan\beta=20$ and $M_R=2$~TeV. The
neutrino data are reproduced by adjusting $\mu_N$.

It is clear that the chargino mixing drops as $v/\MSUSY$ since both
$M_2$ and $\mu$ are approximately linearly dependent on the SUSY
scale. At tree level, the mixing between $\tilde \nu_i$ and $\tilde
\nu^c_j$ is given by $\frac{v}{\sqrt{2}} \big(T^{ji *}_\nu \sin \beta
-\mu Y_\nu^{ji*} \cos \beta + \text{h.c.}\big)$ such that the
left-right mixing matrix entry $Z_{i,3+j}$ also scales approximately
like $v/\MSUSY$. It immediately follows that all mixing effects will
decouple as $\big(v/\MSUSY\big)^2$ since at least two mixing
insertions are necessary. Note that left-left mixing can enhance the
amplitude, but has no impact on the qualitative behavior of the
decoupling with large SUSY masses.
 
\begin{figure}[t]
\center
\includegraphics[width=\linewidth]{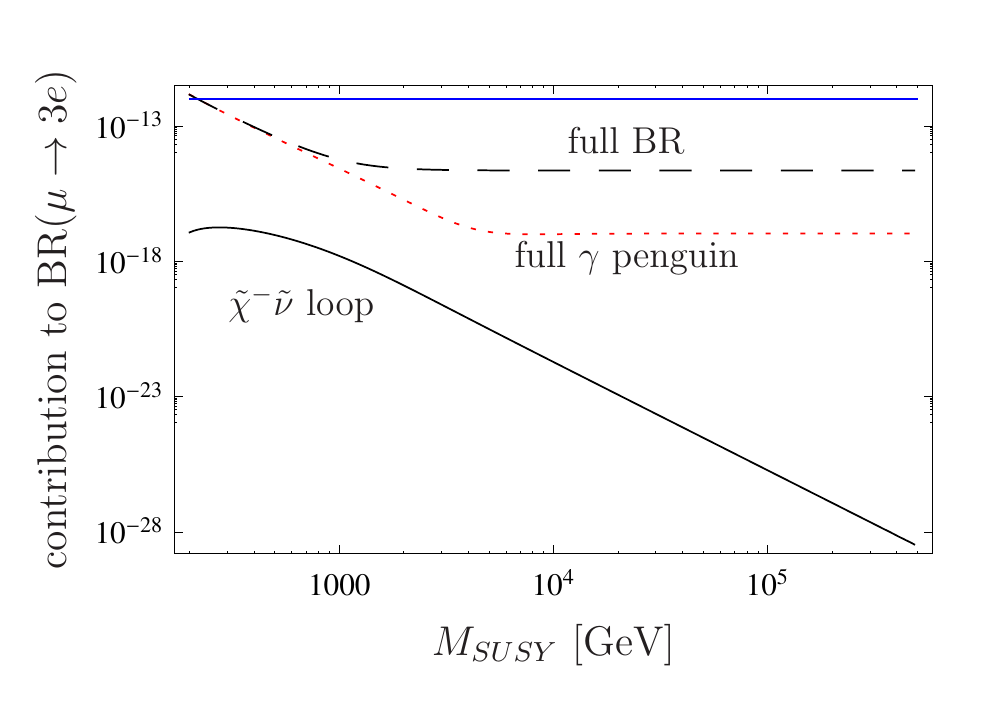}
\caption{Decoupling behavior of BR$(\mu \to 3 e)$ with increasing SUSY
  scale $\MSUSY$ but fixed $M_R=2$~GeV. The dashed black line shows
  BR$(\mu \to 3 e)$, % for $M_R = 2~$TeV and 100~TeV, 
  the dotted red
  line the contribution from the photon penguin only whereas the full
  black line gives the chargino-sneutrino contribution to the $Z$
  penguin.  The other parameters are fixed as explained in the
  text. The blue line shows the experimental upper limit of
  $10^{-12}$ \cite{Bellgardt:1987du}.}
\label{fig:decouplingBehaviorMSUSY}
\end{figure}

\begin{figure}[t]
\center
\includegraphics[width=\linewidth]{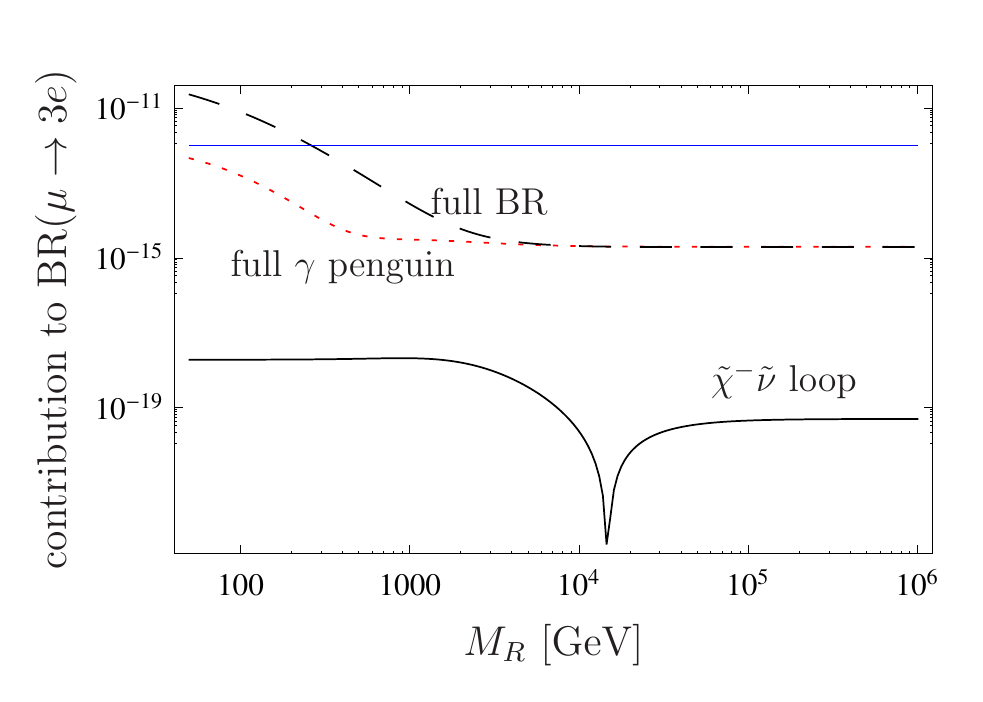}
\caption{Behavior of BR$(\mu \to 3 e)$ with increasing $M_R$ while
  $\MSUSY$ has been fixed to 1~TeV.
The black full line depicts the contribution of the chargino-sneutrino
loop to the $Z$-penguin whereas the red dotted and the black dashed
line show the contribution from the photon penguin and the full
branching ratio. The blue line shows the experimental upper limit of
$10^{-12}$ \cite{Bellgardt:1987du}.}
\label{fig:decouplingBehaviorMR}
\end{figure}

In Fig.~\ref{fig:decouplingBehaviorMSUSY} we show BR($\mu\to 3e)$,
the corresponding photonic contribution as well as the $\tilde
\nu$-$\tilde \chi^-$ contributions as a function of $\MSUSY$. As can
be seen, we obtain the expected decoupling of the SUSY contributions.
Thus, in contrast to, e.g. \cite{Abada:2012cq}, where a nondecoupling
behavior due to a $\big( \MSUSY/M_Z \big)^4$ enhancement of the
$Z$-penguins with respect to the $\gamma$-penguins was found, we do
find the same decoupling behavior of both contributions for large
SUSY scales.  The reason why BR($\mu\to 3e)$ as well as the photonic
contribution are practically constant for $\MSUSY \gsim 4$~TeV is the
$W$-$\nu_i$ non-SUSY contribution. We therefore show in Fig.
\ref{fig:decouplingBehaviorMR} the same quantities as a function of
$M_R$. We can clearly see that one approaches the MSSM limit for $M_R
\gsim 5$~TeV.  In this case the right (s)neutrinos decouple and
only the Higgsino diagram will vanish completely whereas the wino
diagram can still give a large contribution due to possible
chargino and sneutrino left-left mixings.

% added 
An (analytic) comparison to studies independent of
Ref. \cite{Arganda:2005ji}, namely with \cite{Hisano:1995cp,
  Ilakovac:2012sh}, cannot be given here since
Ref. \cite{Hisano:1995cp} did not consider Higgsino contributions to
the $Z$-penguins and Ref. \cite{Ilakovac:2012sh} did not write down
the constant parts of the loop functions (which are responsible for
the earlier found nondecoupling behavior). Nevertheless, the authors
of Ref. \cite{Ilakovac:2012sh} claimed afterwards to agree with our
results \cite{Ilakovac:2014ypa}.

\section{Conclusion: Impact on predictions for LFV in the literature}
\label{sec:conclusion}

We have shown that some recent LFV results in supersymmetric low-scale
seesaw models are based on wrong analytical expressions for the
$Z$-penguins contributing to $\ell_j \to 3 \ell_i$, $\mu-e$ conversion in
nuclei as well as $\tau$ mesonic LFV decays. In fact, this affects not
only the results for inverse seesaw models (or other models with large
superpotential couplings like trilinear $R$-parity violation
\cite{Dreiner:2012mx}), but also studies for models that lead to the
MSSM at low energies~\cite{Arganda:2005ji}.  
However, in the latter case the numerical impact on the LFV violating processes is negligible
since the critical contribution in \EQ{EQ:abada} (induced by light right sneutrinos) is not present. 
In contrast, the
analytical error has a dramatic impact on low-scale seesaw models,
whose phenomenology must be carefully revised. In order to do that, an
independent calculation of all other contributions to the considered
observables is required. Given the interesting new results for the box
contributions to these observables
\cite{Ilakovac:2009jf,Alonso:2012ji,Dinh:2012bp,Ilakovac:2012sh}, it
would be worth confirming by an independent calculation the potential
dominance for $W$-$\nu_R$ boxes in the inverse seesaw in case of low
$M_R$. However, this is beyond the scope of this paper and requires a
complete and independent recalculation of all contributions including
a comparison with previous results. This will be presented elsewhere.

\section*{Acknowledgements}
We thank Martin Hirsch and Maria Jos\'e Herrero for useful
discussions. A.V. is also grateful to Paride Paradisi and Thomas Schwetz
for enlightening discussions. This work has been supported by DFG
research training group 1147 and by DFG project no.\ PO-1337/3-1. A.A., A.V., and C.W.
 aknowledge support from the European Union FP7 ITN
INVISIBLES (Marie Curie Actions, PITN- GA-2011- 289442) and the ANR project CPV-LFV-LHC NT09-508531.

%\bibliographystyle{h-physrev5}
%\bibliography{Decoupling.bib}

\begin{appendix}
\begin{widetext}
\section{Vertices}
\label{appendix:vertices}
Here we provide the vertices for the supersymmetric inverse seesaw model which are relevant for the derivations above.
\begin{align}
C_{iAX(P)}^R = \Gamma^R_{\bar{e}_{{i}}\tilde{\chi}^-_{{A}}\tilde \nu^P_{{X}}}  =  & \,-\frac{i}{\sqrt{2}} \Big(g_2 Z^{P,*}_{X i}   V_{{A 1}}  - \sum_{a=1}^{3}Y^*_{\nu,{a i}} Z^{P,*}_{X 3 + a}  V_{{A 2}} \Big) 
\,,\\ 
C_{iAX(S)}^R = \Gamma^R_{\bar{e}_{{i}}\tilde{\chi}^-_{{A}}\tilde \nu^S_{{X}}}  =  & \,- \frac{1}{\sqrt{2}} \Big(g_2 Z^{S,*}_{X i}   V_{{A 1}}  - \sum_{a=1}^{3}Y^*_{\nu,{a i}} Z^{S,*}_{X 3 + a}   V_{{A 2}} \Big)\,, \\
E^{L(c)}_{BA} =\Gamma^L_{\tilde{\chi}^+_{{B}}\tilde{\chi}^-_{{A}}Z_{{\mu}}}  =  & \,\frac{1}{2} \Big(2 g_2 U^*_{A 1} \cos\theta_W  U_{{B 1}}  + U^*_{A 2} \Big(- g_1 \sin\theta_W   + g_2 \cos\theta_W  \Big)U_{{B 2}} \Big)\,,\\ 
E^{R(c)}_{BA} = \Gamma^R_{\tilde{\chi}^+_{{B}}\tilde{\chi}^-_{{A}}Z_{{\mu}}}  = & \,\frac{1}{2} \Big(2 g_2 V^*_{B 1} \cos\theta_W  V_{{A 1}}  + V^*_{B 2} \Big(- g_1 \sin\theta_W   + g_2 \cos\theta_W  \Big)V_{{A 2}} \Big)\,, \\
Q^{\tilde \nu}_{XY} = \Gamma_{\tilde \nu^P_{{X}}\tilde \nu^S_{{Y}}Z_{{\mu}}}  = & \, -\frac{i}{2} \Big(g_1 \sin\theta_W   + g_2 \cos\theta_W  \Big)\sum_{a=1}^{3}Z^{P,*}_{X a} Z^{S,*}_{Y a}   \,,
\\ 
Z_L^{(\ell)} = \Gamma^L_{\bar{e}_{{i}}e_{{i}}Z_{{\mu}}}  =  & \,\frac{1}{2} \Big(- g_1 \sin\theta_W   + g_2 \cos\theta_W  \Big) \,.
\end{align}

\section{Loop functions}
\label{appendix:loop_functions}
The loop functions in the limit of vanishing external momenta read:
\begin{align}
 B_0 (m_1^2,m_2^2) &= - \log \left(\frac{m_2^2}{Q^2}\right) + \frac{1}{m_2^2 - m_1^2} \Big[m_2^2 - m_1^2 + m_1^2 \log \left(\frac{m_1^2}{m_2^2}\right) \Big]\,,\\
 B_1 (m_1^2,m_2^2) &= -\frac{1}{2} + \frac{1}{2} \log \left(\frac{m_2^2}{Q^2}\right)- \frac{1}{4 (m_1^2 - m_2^2)^2} \Big[ m_1^4 - m_2^4 + 2 m_1^4 \log \left(\frac{m_2^2}{m_1^2}\right)\Big]\,, \\
 C_0 (m_1^2,m_2^2,m_3^2) &= \frac{1}{(m_1^2-m_2^2) (m_3^2-m_1^2) (m_2^2-m_3^2)} \Big[ m_2^2 (m_3^2-m_1^2) \log \left(\frac{m_2^2}{m_1^2}\right)+m_3^2(m_1^2-m_2^2) \log\left(\frac{m_3^2}{m_1^2}\right) \Big]\,,\\
 C_{00} (m_1^2,m_2^2,m_3^2) &= \frac{1}{8(m_1^2-m_2^2) (m_1^2-m_3^2)(m_2^2-m_3^2)} \times \nonumber \\
  \Big[(&m_3^2-m_1^2) \left((m_1^2-m_2^2) (2 \log \left(\frac{m_1^2}{Q^2}\right)-3) (m_2^2-m_3^2)-2 m_2^4 \log\left(\frac{m_2^2}{m_1^2}\right)\right)+2 m_3^4
   (m_2^2-m_1^2) \log \left(\frac{m_3^2}{m_1^2}\right)\Big]\,.
\end{align}
\end{widetext}
\end{appendix}

\end{document}